\documentclass[a4paper,11pt]{article}
\usepackage{pos}
\usepackage{xcolor}

\usepackage{wrapfig}
\usepackage{dblfnote}

\title{Unraveling the Emission Mechanism of the HBL Source Mrk 180 with Multi-Wavelength Data}
 \ShortTitle{Unraveling the Emission Mechanism of Mrk 180}

\author*[a]{Sandeep Kumar Mondal}
\author[b]{Saikat Das}
\author[a]{Nayantara Gupta}

\affiliation[a]{Astronomy \& Astrophysics Group, Raman Research Institute,\\
  C. V. Raman Avenue, Sadashivanagar, Bangalore- 560080, Karnataka, India}

\affiliation[b]{Yukawa Institute for Theoretical Physics, Kyoto University,\\ Kitashirakawa Oiwakecho, Sakyo-ku, Kyoto 606-8502, Japan}

\emailAdd{skmondal@rri.res.in}
\emailAdd{saikat.das@yukawa.kyoto-u.ac.jp}
\emailAdd{nayan@rri.res.in}

\abstract{Markarian (Mrk) 180 is a High frequency-peaked BL Lacertae object or HBL object, located at a redshift of 0.045 and a potential candidate for high-energy cosmic ray acceleration. In this work, we have done a temporal and spectral study using Fermi Large Area Telescope (Fermi-LAT) $\gamma$-ray data, collected over 12.8 years. In the case of the temporal study, the 12.8 years long, 30-day binned, Fermi-LAT $\gamma$-ray light curve does not show any significant enhancement of the flux. To understand the underlying physical mechanism, we focused our study on multi-wavelength spectral analysis. We constructed multi-wavelength spectral energy distribution (MWSED) using Swift X-ray, ultraviolet \& optical, and X-ray Multi-Mirror Mission (XMM-Newton)  data, which have been analysed thoroughly. The SED has been modelled with three different models: (i) pure leptonic scenario and lepto-hadronic scenario where we considered two types of lepto-hadronic interactions (ii) line-of-sight interactions of ultrahigh-energy cosmic rays (UHECR; $E\gtrsim 10^{17}$ eV) with the cosmic background radiation and (iii) interaction between relativistic protons with the cold proton within the blazar jet. In this literature, we have done a detailed comparative study between all these three models. In an earlier study, Mrk 180 was associated with the Telescope Array (TA) hotspot of UHECRs at $E>57$ EeV which motivates us to check whether Mrk 180 can be a source of UHECRs, contributing to the TA hotspot. From our study, we find, for conservative strengths of the extragalactic magnetic field, Mrk 180 is unlikely to be a source of UHECR events.}

\ConferenceLogo{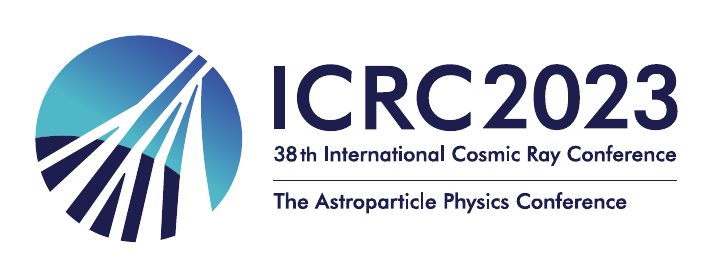}

\FullConference{%
38th International Cosmic Ray Conference (ICRC2023)\\
  26 July - 3 August, 2023\\
  Nagoya, Japan}

\begin{document}
\maketitle

\section{Introduction}
\noindent Active Galactic Nuclei (AGN) is the central core of an active galaxy which emits highly variable radiation ranging from radio to very high-energy (VHE; E$\gtrsim$30 GeV) $\gamma$-rays. The emission from the central core is powered by accretion onto a supermassive black hole (SMBH) which leads to the formation of collimated jets along the direction of the angular momentum. BL Lac is a subclass of AGN whose jet is directed towards the line-of-sight of the observer and has featureless spectra. The low-energy hump of the BL Lac SED is attributed because of the synchrotron radiation from the relativistic leptons. And the most prevalent explanation for the high-energy hump is Inverse Compton (IC) scattering process; in the case of BL Lac, it is considered as synchrotron self-Compton (SSC). The presence of VHE $\gamma$-rays is also possible because of the radiation from photohadronic (p$\gamma$) or hadronuclear (pp) interactions of the cosmic rays with the ambient medium in the emission zone or blazar jet or proton synchrotron emission.\\
Mrk 180 is a BL Lac object, located at a redshift= 0.045, R.A.= 174.11008 deg, Decl.= 70.1575 deg, discovered by a Swiss-origin astronomer Fritz Zwicky. In March 2006, VHE $\gamma$-ray emission was detected for the first time \citep{2006outburst} from this source, triggered by an optical burst. This source was monitored by several telescopes e.g. Fermi-LAT, Swift, Major Atmospheric Gamma Imaging Cherenkov Telescope (MAGIC), XMM-Newton, Monitoring of jets in Active Galactic Nuclei with VLBA Experiments (MOJAVE), KVA, ASM. An earlier study \citep{He} identified Mrk 180 as a possible source of UHECRs in the context of explaining the origin of the TA hotspot at $E>57$ EeV.  Motivated by this earlier study, we carry out a comprehensive study of Mrk 180 to ascertain the underlying mechanism of high-energy $\gamma$-ray emission and whether it can be the source of UHECRs beyond 57 EeV contributing to the TA hotspot. In this work, we have mainly carried out a temporal and spectral study of Mrk 180.

\section{\label{sec:data} Data Analysis}
\begin{wrapfigure}{r}{0.5\textwidth}
\centering
    \includegraphics[width=0.495\textwidth, height=0.25\textwidth]{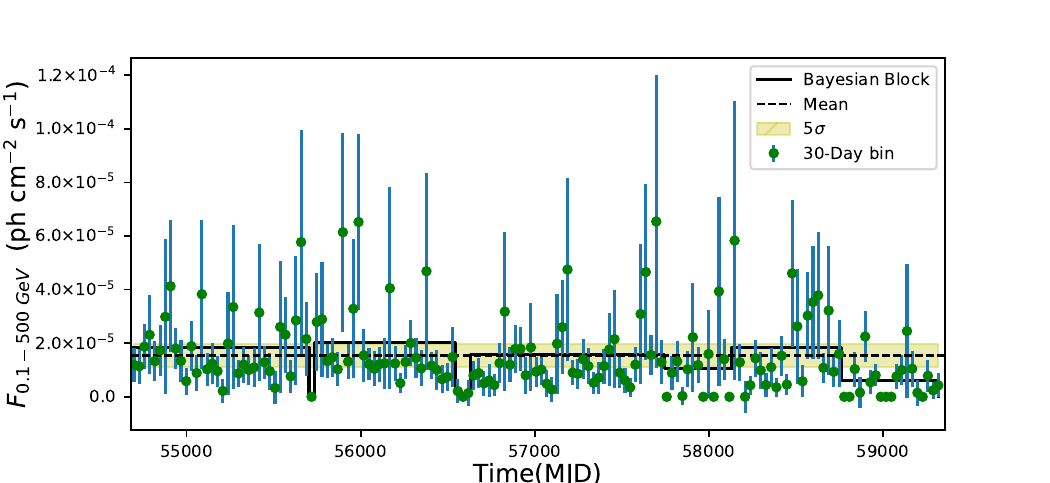}
    \caption{Application of Bayesian Block Method on Fermi-LAT $\gamma$-ray light curve of Mrk 180 (MJD 54682.65- 59355.67)}
    \label{fig:Mrk180_LC}
\end{wrapfigure}

\noindent \textbf{Fermi-LAT:}
\noindent The Fermi-LAT is an imaging, pair-conversion, wide-field-of-view, high-energy $\gamma$-ray telescope that can detect photons of energy 20 MeV to more than 300 GeV, with a field of view of 2.4 sr \citep{Atwood_2009}. The LAT is Fermi's primary instrument. For this work, we have extracted the Pass 8 Fermi-LAT $\gamma$-ray data of Mrk 180 from Fermi Science Support Center (FSSC) data server from August 2008 to May 2021, almost 12.8 years. We have analyzed this data with Fermipy (v1.0.1; \cite{Fermipy}) tool following the standard data analysis procedure. The long-term Fermi-LAT $\gamma$-ray light curve has been shown in Fig.~\ref{fig:Mrk180_LC} and the SED, which is used to construct the MWSED shown in Fig.~\ref{fig:Pure_LP_MWSED_Lp_UHECR_MWSED} and ~\ref{fig:LP_PP_MWSED}.

\noindent \textbf{SWIFT XRT and UVOT:}
\noindent  Neil Gehrels Swift observatory is a multi-wavelength space-based observatory with three instruments onboard: Burst Alert Telescope (BAT), X-Ray Telescope (XRT) and Ultraviolet and Optical Telescope (UVOT)  \citep{Burrows}. We collected all the XRT and UVOT data of Mrk 180 over the period from August 2008 to May 2021 (44 observations). The standard data reduction procedure has been followed for the analysis e.g. source \& background region selection etc. The final X-ray SED has been obtained after being modelled by xspec (v12.11.0; \citep{xspec}). Similarly, we have obtained UVOT SED points in all six filters, considering the galactic absorption. We have used the following extinction coefficient values corresponding to different Swift-UVOT wavebands which have been obtained from the python module ‘extinction’ \footnote{\url{https://extinction.readthedocs.io/en/latest/\#}}; U: 0.05584, V: 0.03460, B: 0.04603, UVW1: 0.07462, UVM2: 0.10383, UVW2: 0.09176.


\noindent \textbf{XMM-Newton X-ray Data Analysis:}
\noindent XMM-Newton is a space-borne X-ray observatory, consisting of three imaging X-ray cameras (European Photon Imaging Camera or EPIC), two grating X-ray spectrometers (Reflection Grating Spectrometer or RGS) and one optical monitor (OM). The three EPIC cameras are the primary instrument aboard XMM-Newton; out of the three, two of them are MOS-CCD cameras and the remaining one is the pn-CCD camera. From the data archive of XMM-Newton \footnote{\url{http://nxsa.esac.esa.int/nxsa-web/##search}}, we found two observations for Mrk 180: 0094170101 and 0094170301 of 20 ks and 8 ks respectively. We have followed standard data reduction procedure \footnote{\url{https://www.cosmos.esa.int/web/xmm-newton/sas-threads}} to extract the SED. Finally, we got SED data points from the both MOS and pn detector. Thereafter, we used xspec  (v12.11.0; \cite{xspec}) to model these spectra. We also analyzed OM image mode data. Following the same data reduction procedure for further analysis. The first observation 0094170101, contains single data corresponding to the u-band, which is insufficient for further analysis whereas, the second observation 0094170301 does not contain any image file for further study. So, our multi-wavelength data does not contain any XMM-Newton OM data.

\noindent \textbf{Archival Data : MOJAVE, MAGIC \& SSDC:}
\noindent We have used archival data from MOJAVE, MAGIC and SSDC. MOJAVE is a long-term program to monitor radio brightness and polarization variation in jets associated with active galaxies visible in the northern sky. We have collected MOJAVE data for Mrk 180, which consists of seven observations, from the MOJAVE/2cm Survey Data Archive \footnote{\url{https://www.cv.nrao.edu/MOJAVE/sourcepages/1133+704.shtml}}. This research has made use of data from the MOJAVE database that is maintained by the MOJAVE team \citep{mojaveteam}.
 MAGIC is a system of two Imaging Atmospheric Cherenkov Telescopes (IACT), situated on the Canary Island of La Palma, which can detect $\gamma$-rays of energy between 30 GeV to 100 TeV. VHE $\gamma$-rays from Mrk 180 were detected during an optical outburst in 2006 \citep{2006outburst}. We have used that data from \footnote{\url{http://vobs.magic.pic.es/fits/\#database}} for our study.
And lastly, we have collected the data from SSDC SED builder \footnote{\url{https://tools.ssdc.asi.it/SED/}} and shown it with grey squares in the multi-wavelength SEDs (Fig.~\ref{fig:Pure_LP_MWSED_Lp_UHECR_MWSED} and ~\ref{fig:LP_PP_MWSED}).


\section{\label{subsec:light_curve} Long-term Fermi-LAT gamma-Ray Light curve}

\noindent The 12.8 years long, 30-day binned Fermi-LAT $\gamma$-ray light curve Fig.~\ref{fig:Mrk180_LC} does not show any significant variation in the $\gamma$-ray flux. There are a few data points with high $\gamma$-ray flux with large error bars, so further temporal study is not feasible in this case.



\section{\label{sec:sed_model}Multi-Wavelength SED Modeling}

\noindent  We have constructed the MWSED with analyzed and archival data (as mentioned in sec.~\ref{sec:data}) which covers radio to $\gamma$-ray wavebands. The MWSED has been modelled with three different models, discussed below.

\noindent \textbf{Pure Leptonic Modeling:}
\noindent We have considered a spherical emission region of radius $R$ within the jet, moving with a Doppler factor $\delta_D$, where relativistic electrons and positrons accelerated in the jet, lose energy through synchrotron radiation in a steady and uniform magnetic field $B$, and also by SSC emission. From the maximum likelihood analysis of Fermi-LAT data, a log-parabola injection was found to best fit the data. Following, \cite{massaro} we have used the log-parabolic spectrum of the injected electrons in the blob to explain the MWSED of Mrk 180. 
\noindent We have used an open-source code, `GAMERA' \cite{Hahn2016} to model the MWSED. We consider a constant escape of the electrons from the emission region over the dynamical timescale. We find that the time-evolved electron spectrum reaches the steady state after nearly 100 days, and this spectrum has been used in this work.

\begin{figure*}[h]
    \centering
    \includegraphics[width=0.495\textwidth, height=0.45\textwidth]{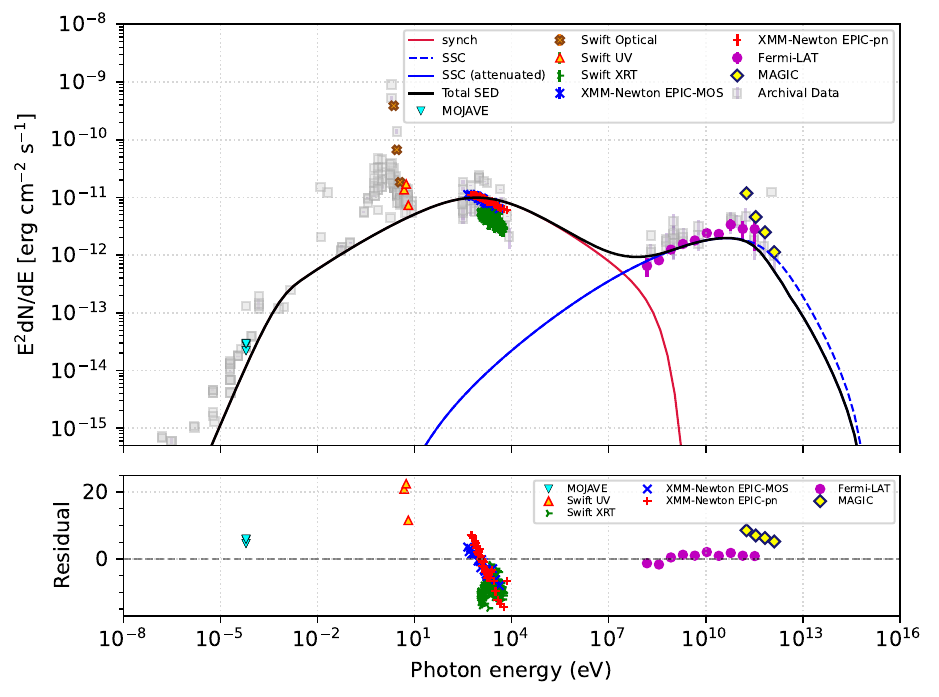}
     \includegraphics[width=0.495\textwidth, height=0.45\textwidth]{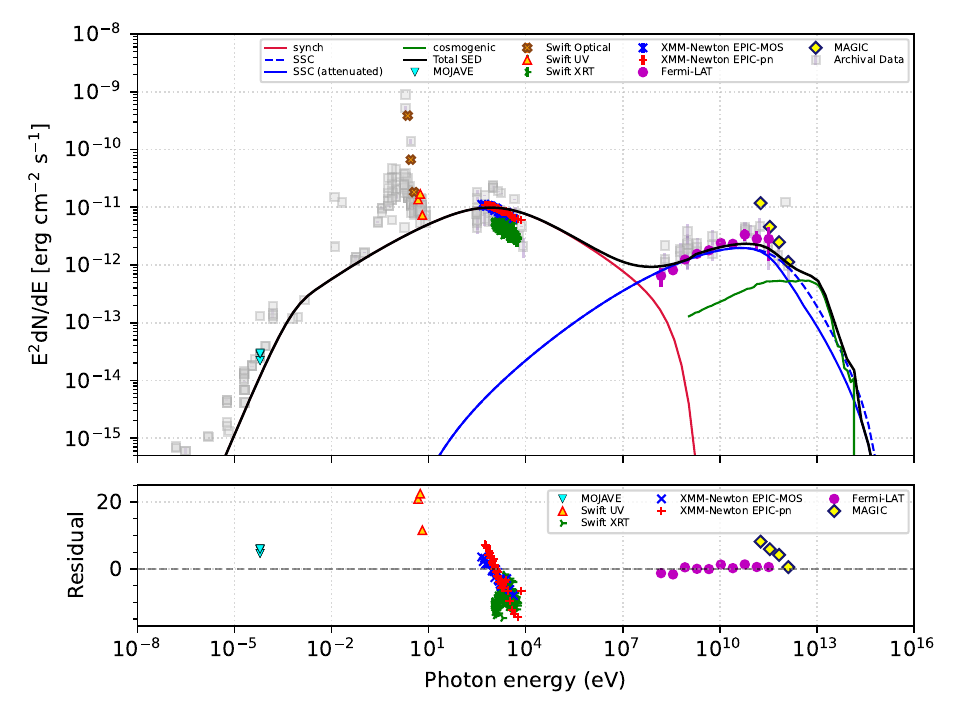}
    \caption{Left: (a) Pure leptonic modeling of MWSED of Mrk 180; Right: (b) Leptonic+ hadronic (UHECR) modeling of MWSED of Mrk 180; with residual plot corresponding to the modeling. The data color codes are mentioned in the plots.}
    \label{fig:Pure_LP_MWSED_Lp_UHECR_MWSED}
      
\end{figure*}



\noindent \textbf{UHECR Interactions:}
\noindent We have assumed a power-law injection of the protons into the interstellar medium (ISM) within an energy range of 0.1-100 EeV. The ultra-high energy protons escape from the emission region and propagate through the extra-galactic medium interacting with CMB and EBL photons.  In this process, electrons, positrons, $\gamma$-rays, and neutrinos are produced through $\Delta$-resonance and Bethe-Heitler pair production. The neutral pions decay to $\gamma$ photons ($\pi^\circ \rightarrow \gamma \gamma$) and the charged pions decay to neutrino ($\pi^+ \rightarrow \mu^+ + \nu_\mu \rightarrow e^+ + \nu_e + \bar{\nu}_\mu + \nu_\mu $). The resulting cosmogenic neutrinos propagate undeflected by magnetic fields and unattenuated by interaction with other particles.\\
\noindent The secondary $e^\pm$, $\gamma$-rays initiate electromagnetic (EM) cascade by undergoing pair production, inverse-Compton upscattering of the background photons, and synchrotron radiation in the extragalactic magnetic field (EGMF). The resulting spectrum extends down to GeV energies.\\
\noindent We have used the publicly available simulation framework, \textsc{CRPropa 3} \citep{batista2016, batista2022} to propagate UHECR protons (for simplicity we have considered only protons) from their source to the observer. The secondary EM particles are propagated in the CRPropa simulation chain, using a value of EM thinning $\eta=0.6$.

\noindent \textbf{ pp Interactions:}
\noindent When the relativistic protons have much lower energy than UHECRs and they interact with the cold protons within the emission region as they are trapped in the magnetic field of the emission region. The proton-proton interactions result in the production of neutral and charged pions. These pions decay into secondary particles e.g. electrons/ positrons, neutrinos and $\gamma$-rays.
\begin{wrapfigure}{r}{0.5\textwidth}
  \centering
    \includegraphics[width=0.52\textwidth,height=0.45\textwidth]{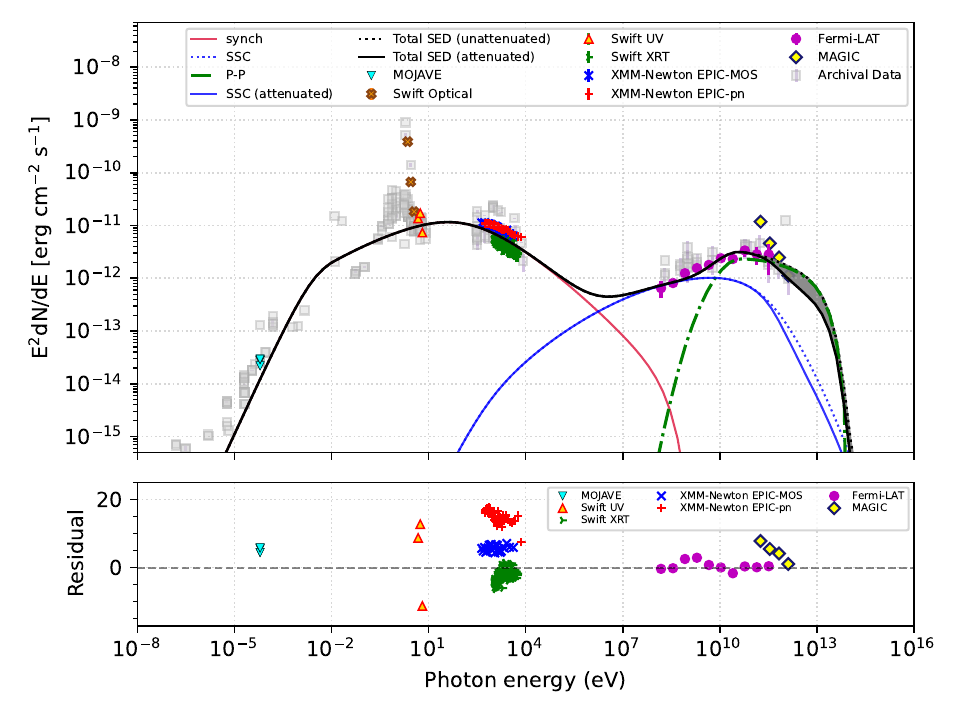}
        \caption{Leptonic+ hadronic ($pp$) modeling of MWSED of Mrk 180 and residual plot corresponding to this modeling; the grey-shaded region denotes the difference between the attenuated and unattenuated regions of the total SED.}
    \label{fig:LP_PP_MWSED}
\end{wrapfigure}
\noindent We have considered a power-law proton injection spectrum within the emission region, with a proton spectral index $\alpha_p$  and energy range 10- 10$^4$ GeV with the publicly available code \textsc{Gamera} for the time-independent $pp$ modelling. We have balanced the total charge in the emission region to determine the total number of protons. The $\gamma$-ray spectrum produced in $pp$ interactions has been corrected for internal absorption by the lower energy photons inside the blob, and also for absorption by the EBL.


\subsection{\label{subsec:power}Jet Power}
\noindent We have calculated the total kinematic jet power using the following equation
$ P^k_{\rm tot}= P_e+ P_B + P_p =\pi R^2 \Gamma^2 c (u'_e+u'_p+u'_B)$
where P$^k_{\rm tot}$ is the kinematic jet power, $\Gamma$ is the bulk Lorentz factor; $u'_e$, $u'_p$ and $u'_B$ are the energy densities of the relativistic electrons (and positrons) and protons and magnetic field respectively in the comoving jet frame \citep{bb2019, bb2020}. The primed and unprimed notations denote quantities in the comoving jet frame and the AGN frame, respectively. We have maintained the charge neutrality condition in the jet. If we add the jet power of cold protons the luminosity budget in the proton-proton interaction model exceeds the Eddington luminosity as discussed in \citep{bb2019, bb2020}. We calculated the Eddington luminosity of Mrk 180 from the mass of Mrk 180, which is 5.06-6.51$\times 10^{46}$ erg/s. We compare only the kinematic jet power to the Eddington luminosity as it has been done in earlier papers.

\section{\label{sec:results}Results}
\noindent The 12.8 years long 30-day binned Fermi-LAT $\gamma$-ray light curve (Fig.~\ref{fig:Mrk180_LC}) of Mrk 180 does not show any significant flaring throughout this time, also the error bars of the high-energy $\gamma$-ray data points are large.
\noindent To know about the physical processes, we further investigate the long-term MWSED of Mrk 180. The MWSED shows the double hump structure, which has been modelled with GAMERA; considering a simple one-zone spherical emission region within the jet. In Fig.~\ref{fig:Pure_LP_MWSED_Lp_UHECR_MWSED} and ~\ref{fig:LP_PP_MWSED}, we have shown the MWSEDs fitted with different models e.g. pure leptonic, lepto-hadronic. Also, we have shown the residual (Data-Model/error) plot corresponding to the fit to each model in Fig.~\ref{fig:Pure_LP_MWSED_Lp_UHECR_MWSED} and ~\ref{fig:LP_PP_MWSED}.

\noindent In the case of a pure leptonic model, the first hump is produced due to the synchrotron radiation of the relativistic electrons, and the second hump is produced due to the up-scattering of the synchrotron photons by the relativistic electrons. From Fig.~\ref{fig:Pure_LP_MWSED_Lp_UHECR_MWSED} (a) we can see that the SED from the pure leptonic model cannot fit the Swift UV data points. The slope of the observed X-ray and the $\gamma$-ray data points cannot be explained with the slope of the theoretical SED; it poorly fits the $\gamma$-ray data points, also the highest energy $\gamma$-ray data point cannot be fitted with this model. The residual plot corresponding to the pure leptonic model shows this model poorly fits the Swift UV data, X-ray, and MAGIC data. So this model cannot explain the MWSED at all. 

\begin{figure*}[h]
\small
    \centering
    \includegraphics[width=0.32\textwidth]{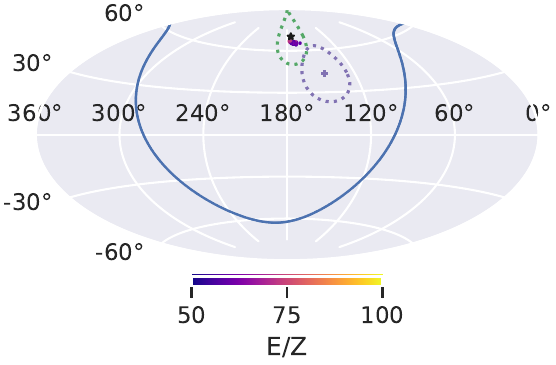}%
    \includegraphics[width=0.32\textwidth]{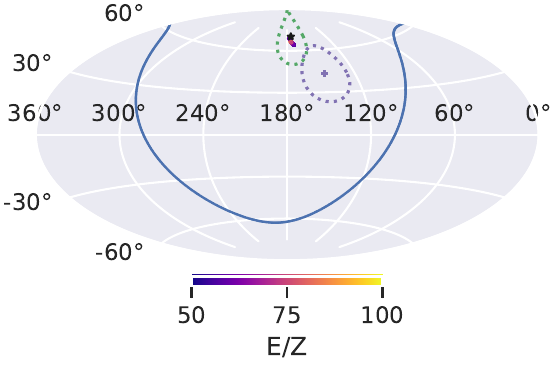}%
    \includegraphics[width=0.32\textwidth]{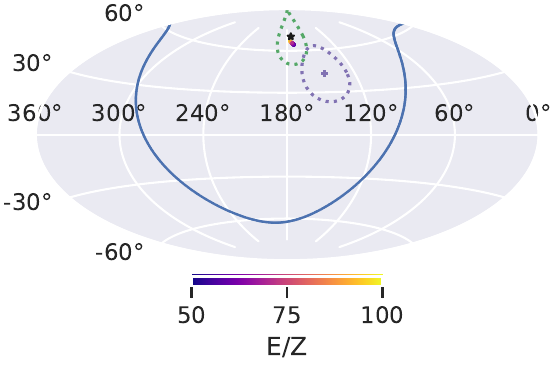}
    \caption{Arrival direction of UHECRs at $E>57$ EeV from Mrk 180 to Earth. The blue line shows the Galactic plane. The purple point and the purple dotted curve show the TA hotspot center and the 20$^\circ$ region around it. Similarly, the green dotted curve shows the $20^\circ$ region around Mrk 180. The color bar indicates the energy per nucleon (E/z) of the observed events. From left, the figures correspond to (a) pure proton injection and B$_{\rm rms }\approx10^{-3}$ nG; (b) pure proton injection and B$_{\rm rms}\approx10^{-5}$ nG; (c) Fe injection and B$_{\rm rms}\approx10^{-5}$ nG}. 
    \label{fig:aniso}
\end{figure*}

 \noindent For the improvement of the fit, particularly at the VHE $\gamma$-ray regime, we check the fit with lepto-hadronic models (Fig.~\ref{fig:Pure_LP_MWSED_Lp_UHECR_MWSED} (b) and Fig.~\ref{fig:LP_PP_MWSED}). 
In the case of UHECRs, the escape of protons from the blazar jet can dominate over the energy loss inside the blazar jet. We have considered power-law injection of protons into ISM between 0.1-100 EeV with proton spectral index ($\alpha_p)$ 2.2. In this model, we consider the three-dimensional propagation of UHECRs to calculate the fraction of them that survives within $0.1^\circ$ degrees of initial emission direction. We consider a random turbulent EGMF given by a Kolmogorov power spectrum and an RMS field strength of B$_{\rm rms}\approx10^{-5}$ nG and a coherence length of 0.5 Mpc. We consider all these factors to calculate the $\gamma$-rays reaching the observer from the direction of the BL Lac. Fig.~\ref{fig:Pure_LP_MWSED_Lp_UHECR_MWSED} (b) is the resulting fit corresponding to this model. The green curve indicates the spectrum of cosmogenic photons. In this case, the highest energy MAGIC data point can be fitted, but the fit to the X-ray data points has not improved. Moreover, the Swift UV data cannot be fitted well with this model. The residual plot corresponding to this model looks almost the same as that of the pure leptonic model between 10$^{-5}$- 10$^{11}$ eV, except for the MAGIC data points.\\
\noindent We subsequently consider the $pp$ interactions within the jet. As explained in sec. \ref{sec:sed_model}, the relativistically accelerated protons interact with cold protons and produced neutral and charged pions which decay into photons, leptons, and neutrinos. The cold proton density is assumed to be n$_{\rm H}$=1.2$\times10^6$ cm$^{-3}$. Fig.~\ref{fig:LP_PP_MWSED} shows improvement in both SED and the residuals. The SED fits the Swift UV data points and matches the slope of the X-ray data and the $\gamma$-ray data. We have not shown the residuals for the Swift Optical data points, as they cannot be fitted with any of these models. Most of the $\gamma$-ray data points can be fitted in this model.\\
\noindent The total kinematic jet power corresponding to each model is less than the Eddington luminosity of Mrk 180, which has been mentioned in Table~\ref{tab:GAMERA_Fitting_param}. Also, the best-fitted parameter values corresponding to each model are listed in the same table. \\
\noindent To check the association of Mrk 180 as a UHECR source to the TA hotspot at $E>57$ EeV, we propagate UHECRs from the source to the Earth in a random turbulent magnetic field given by the Kolmogorov power spectrum. We consider three different combinations of the RMS value of the EGMF (B$_{\rm rms}$) and composition ($^1$H and $^{56}$Fe) at the source as shown in Fig.~\ref{fig:aniso}. The turbulence correlation length of the EGMF is taken to be ~0.5 Mpc. The Galactic magnetic field model (GMF) is considered to be the one given in Jansson \& Farrar \cite{JF}. We inject cosmic rays with a generic power-law spectrum given by $dN/dE\sim E^{-2}$ and perform three-dimensional simulations including both GMF and EGMF in \textsc{CRPropa 3} \citep{batista2016,batista2022}. \\
After considering different B$_{\rm rms}$ and compositions,  we found that for conservative strength of EGMF, the contribution of this source to the TA hotspot is disfavored. Thus, Mrk 180 may not be a plausible UHECR source for explaining the TA hotspot.

\section{\label{sec:discussion}Discussions \& Conclusion}

\noindent  There are no significant flux variation in the long-term Fermi-LAT $\gamma$-ray light curve (Fig.~\ref{fig:Mrk180_LC}). Also, the error bars of the high-energy $\gamma$-ray data points are large to carry on a detailed temporal study on this source. Hence a more detailed analysis of the light curve cannot give us any useful information.\\
\noindent We modelled the MWSED with GAMERA to explore the underlying emission mechanism of Mrk 180. It is found that a single-zone pure leptonic model cannot explain the multi-wavelength spectrum of Mrk 180 properly. We considered single-zone lepto-hadronic models to obtain better fits to the data. The residuals of the three models are compared and the $pp$ interaction model is found to give a better fit to the multi-wavelength data compared to the other two models; however, more observational data is necessary to explain the radiation mechanisms in Mrk 180, as our results show large values of residuals in all the cases. We look forward to future multi-wavelength campaigns to cover all the frequencies over a long period to monitor this source more closely to give a definitive conclusion.
  
\par
\noindent \cite{He} calculated the probability associated with some sources to be the contributors to the TA hotspot, Mrk 180 is one of them. It is important to know the role of Mrk 180 as a UHECR accelerator, and whether it can generate events above 57 EeV. In our study we found that Mrk 180 is disfavoured as a source of the UHECR events contributing to the TA hotspot considering conservative strength of magnetic field. In future, with more observational data it would be interesting to study the association of Mrk 180 with the TA hotspot.

\begin{table*}
\small
\caption{Results of multi-wavelength SED modeling shown in the Fig. ~\ref{fig:Pure_LP_MWSED_Lp_UHECR_MWSED} (a), \ref{fig:Pure_LP_MWSED_Lp_UHECR_MWSED} (b) and ~\ref{fig:LP_PP_MWSED}}
\centering
\begin{tabular}{cccc rrrrr}   
\hline\hline
 Parameters & & Values &  &\\
\hline
& Pure-leptonic  & Leptonic+ hadronic (UHECR) & Leptonic+ hadronic ($pp$) \\
& model & model & model \\
\hline
 $\alpha$ &  2.2 & 2.2& 2.2&\\
 $\beta$ & 0.06& 0.06 & 0.10 &\\
 B &  0.10 G & 0.10 G & 0.10 G &\\
 R &  8.0$\times10^{15}$ cm & 8.0$\times10^{15}$ cm& 1.8$\times10^{16}$ cm& \\
 $\delta_D$ &  20 & 20& 20& \\
 $\gamma_{\rm min}$  & 1.0$\times10^2$ & 1.0$\times10^2$& 2.5$\times10^2$& \\
 $\gamma_{\rm max}$  & 9.0$\times10^7$ & 9.0$\times10^7$ & 9.0$\times10^7$& \\
 $\alpha_p$ &-- & 2.2 & 2.2\\
 P$^k_{\rm tot}$ &  2.7$\times10^{43}$ erg/s & 2.9 $\times10^{43}$ erg/s & $1.0\times10^{45}$ erg/s& \\
\hline

\end{tabular}
\label{tab:GAMERA_Fitting_param}
\end{table*}

%
%
%


\newpage
\begin{thebibliography}{99}

\bibitem{2006outburst}
Albert, J., Aliu, E., Anderhub, H., et al. 2006, ApJ, 648, L105

\bibitem{He}
He, H.-N., Kusenko, A., Nagataki, S., et al. 2016, PhRvD, 93, 043011

\bibitem{Atwood_2009}
Atwood, W. B., Abdo, A. A., Ackermann, M., et al. 2009, ApJ, 697, 1071

\bibitem{Fermipy}
Wood, M., Caputo, R., Charles, E., et al. 2017, ICRC, 35, 824

\bibitem{Burrows}
Burrows, D. N., HPhysRevD.99.103006, PhysRevD.101.063024ill, J. E., Nousek, J. A., et al. 2005, SSRv, 120, 165

\bibitem{xspec}
Arnaud, K. A. 1996, in ASP Conf. Ser. 101, Astronomical Data Analysis
Software and Systems V, ed. G. H. Jacoby \& J. Barnes (San Francisco, CA:
ASP), 17

\bibitem{mojaveteam}
Lister, M. L., Aller, M. F., Aller, H. D., et al. 2018, ApJS, 234, 12

\bibitem{massaro}
Massaro, E., Perri, M., Giommi, P., \& Nesci, R. 2004, A\&A, 413, 489

\bibitem{Hahn2016}
Hahn, J. 2016, ICRC (The Hague), 34, 917

\bibitem{batista2016}
Alves Batista, R., Dundovic, A., Erdmann, M., et al. 2016, JCAP, 2016, 038

\bibitem{batista2022}
Alves Batista, R., Becker Tjus, J., Dörner, J., et al. 2022, JCAP, 2022, 035

\bibitem{bb2019}
Banik, P., \& Bhadra, A. 2019, PhRvD, 99, 103006

\bibitem{bb2020}
Banik, P., Bhadra, A., Pandey, M., \& Majumdar, D. 2020, PhRvD, 101,
063024

\bibitem{JF}
Jansson, R., \& Farrar, G. R. 2012, ApJ, 757, 14












\end{thebibliography}
\end{document}